\documentclass[aps,preprintnumbers,floatfix,nofootinbib,preprint,superscriptaddress]{revtex4}

\usepackage{amsmath}
\usepackage{graphicx}

\DeclareMathOperator{\erf}{erf}

\DeclareMathOperator{\CDF}{CDF}
\DeclareMathOperator{\odds}{Odds}

\newcommand{\NOvA}{NO$\nu$A}
\newcommand{\barH}{{\bar H}}

\newcommand{\Li}{\mathcal{L}}

\begin{document}

\title{On the Bayesian approach to neutrino mass ordering}

\author{Mattias Blennow}
\email{emb@kth.se}
\affiliation{Department of Theoretical Physics, School of Engineering Sciences, KTH Royal Institute of Technology, AlbaNova University Center, 106 91 Stockholm, Sweden}

\begin{abstract}
We study the framework of Bayesian statistics for analyzing the capabilities and results of future experiments looking to solve the issue of the neutrino mass ordering. Starting from the general scenario, we then give examples of the procedure for experiments with Gaussian and non-Gaussian distributions for the indicator. We describe in detail what can and cannot be said about the neutrino mass ordering and a future experiment's capabilities to determine it. Finally, we briefly comment on the application to other binary measurements, such as the determination of the octant of $\theta_{23}$.
\end{abstract}

\maketitle

\section{Introduction}

With the lepton mixing angle $\theta_{13}$ being discovered to be large~\cite{An:2012eh,An:2012bu,Ahn:2012nd,Abe:2012tg}, there has been a surge of interest in the possibility of determining whether the neutrino mass ordering is normal (NO) or inverted (IO) in the next or next-to-next generation of neutrino experiments. Many studies of the capabilities for doing this in different types of experiments have been performed, including atmospheric~\cite{Banuls:2001zn,TabarellideFatis:2002ni,Bernabeu:2003yp,PalomaresRuiz:2004tk,Indumathi:2004kd,Petcov:2005rv,Samanta:2006sj,Kopp:2007ai,Gandhi:2007td,Donini:2007qt,Mena:2008rh,Gandhi:2008zs,Samanta:2009qw,FernandezMartinez:2010am,Blennow:2012gj,Akhmedov:2012ah,Ghosh:2012px,Agarwalla:2012uj,Franco:2013in,Ribordy:2013xea,Winter:2013ema,Ghosh:2013mga,Blennow:2013vta,Ge:2013zua}, reactor~\cite{Petcov:2001sy,Schonert:2002ep,Choubey:2003qx,Nunokawa:2005nx,Learned:2006wy,Zhan:2008id,Zhan:2009rs,Ghoshal:2010wt,Qian:2012xh,Ciuffoli:2012iz,Ghoshal:2012ju,Ge:2012wj,Ciuffoli:2012bs,Ciuffoli:2013ep,Li:2013zyd,Kettell:2013eos,Capozzi:2013psa,Blennow:2013vta,Ciuffoli:2013pla}, and long baseline~\cite{Cervera:2000kp,Barger:2000cp,Nunokawa:2005nx,BurguetCastell:2002qx,Huber:2002rs,Minakata:2003ca,Donini:2007qt,Coloma:2007nn,Huber:2009cw,Agarwalla:2011hh,Coloma:2011pg,Prakash:2012az,Coloma:2012ma,Coloma:2012ut,Dusini:2012vc,Blennow:2012gj,Agarwalla:2012bv,Agarwalla:2013hma,Barger:2013rha,Qian:2013nhp,Messier:2013sfa,Ghosh:2013pfa,Bass:2013hla} neutrino experiments. Most of these studies take an approach where the neutrino mass ordering is determined in a frequentist manner with the typical square root of the test statistic used as a measure of the sensitivity as if the distribution of the test statistic was a $\chi^2$ distribution with one degree of freedom. The issue of how to deal with the fact that Wilks' theorem~\cite{Wilks} does not apply to a binary measurement has also been dealt with in different ways in several studies~\cite{Qian:2012zn,Ge:2012wj,Ciuffoli:2013rza,Capozzi:2013psa}. The statistical analysis of the ordering measurement can be performed in either a frequentist or a Bayesian statistics setting. Although the Bayesian method of model selection is ideally suited for the task, the neutrino community is traditionally frequentist and more accustomed to interpreting frequentist results. Thus, in choosing which approach to take, these two facts have to be weighted against each other. It should be mentioned that the frequentist analysis can lead to some results that may be considered unappealing, such as the rejection of both hierarchies at high confidence, there is nothing wrong in performing it as long as proper care is taken in interpreting the results. In this text we will concentrate on how to correctly perform the Bayesian analysis, both as a method of interpreting actual results as well as for judging the capabilities of future experiments. The frequentist approach will be discussed elsewhere~\cite{Blennow:2013oma}.

The remainder of this paper has been organized as follows: We start by quickly reviewing the Bayesian approach to model selection in Sec.~\ref{sec:bayesian}. In Sec.~\ref{sec:twomodels} we then specialize to the situation where we have two models that we wish to compare, including definitions that we will use later as well as an analytical treatment of the case where a Gaussian approximation is valid. We continue by giving an example of a non-Gaussian situation in Sec.~\ref{sec:nongaussian}, where we also briefly discuss a semi-analytic approach for the case when the distributions are essentially sums of different Gaussian distributions. In this section we also discuss the prior dependence of Bayesian analyses and its implications for decision making in future neutrino oscillation experiments. Finally, in Sec.~\ref{sec:summary}, we discuss and summarize our results.

\section{Bayesian model selection}
\label{sec:bayesian}

Unlike frequentist statistics, which are only concerned with how probable outcomes were given a hypothesis, Bayesian statistics deal with our degree of belief in a hypothesis. This has both advantages and disadvantages, the main disadvantage being the introduction of a final result which may depend on the prior knowledge that is inserted into the analysis. However, in some cases it may be of interest to actually consider the prior and changes to it as valuable tools at the design level of an experiment (see, e.g., Ref.~\cite{Blennow:2013swa}). Bayesian model selection is performed as follows: Consider a situation where we have to select among several different hypotheses $H_i$, all of which are mutually exclusive. Before any experiment is performed, we assign a \emph{prior} $\pi_i = P(H_i)$ to each hypothesis. These priors are quantifications of our degree of belief in the different hypotheses before having any experimental input and thus are typically chosen such that $\pi_i = \pi_j$ unless some arguments can be applied for having greater belief in one hypothesis than in the others (or if some other experiment has already provided an indication that this is the case). In general
\begin{equation}
 \sum_i \pi_i \leq 1,
\end{equation}
where the equality only holds if we are absolutely convinced that one of the hypotheses is true. Once we have performed an experiment, we wish to update our degree of belief in the different hypotheses by computing how likely each hypothesis is, this quantity is the \emph{posterior probability} $P(H_i;t)$, i.e., the probability that $H_i$ is true given an observation $t$. What we are really interested in are the \emph{posterior odds}
\begin{equation}
 \odds(H_i,H_j;t) = \frac{P(H_i;t)}{P(H_j;t)} \equiv \frac{p_{H_i}}{p_{H_j}}.
\end{equation}
Central to Bayesian statistics, we now make use of Bayes' theorem
\begin{equation}
 P(A,B) = P(A;B) P(B) = P(B;A) P(A) \quad \Rightarrow \quad p_{H_i} \propto \mathcal L_{H_i}(t) \pi_i
\end{equation}
to rewrite the odds as
\begin{equation}
 \odds(H_i,H_j;t) = \frac{\mathcal L_{H_i}(t)}{\mathcal L_{H_j}(t)} \frac{\pi_i}{\pi_j}.
\end{equation}
Here, $\mathcal L_{H_i}(t)$ is the likelihood of producing the observation $t$ if the hypothesis $H_i$ is true. Thus, the posterior odds are simply updates of the prior odds such that each hypothesis is weighted by how likely the outcome was given the hypothesis.

In many situations, the hypotheses that are being treated are not simple, but rather composed of a general model with one or several model parameters~$\theta$. In these cases, we also need to assign a parameter prior $\pi_{H_i}(\theta_i)$ quantifying how likely each realization of parameters are. The likelihood of the data is then in general dependent on the parameters $\theta_i$. However, the model likelihood is easily computable by weighting the likelihoods with the parameter prior
\begin{equation}
 \mathcal L_{H_i}(t) = \int \pi_{H_i}(\theta_i) \mathcal L_{H_i}(t;\theta_i)\ d\theta_i. \label{eq:likelihood}
\end{equation}

In general, the Bayesian approach would consider the data as is without referencing a test statistic. However, by only considering a single observable $T$, we are going to simplify the analysis significantly and this is just what we will do below. Since we are not dealing with the frequentist notion of hypothesis testing, we will refrain from calling this $T$ a test statistic and instead refer to it as an \emph{indicator}. This indicator will be well suited for telling two hypotheses, $H_i$ and $H_j$, apart if its distribution under the different hypotheses differ significantly. Keeping this in mind, we now concentrate on the case at hand, where we want to distinguish two given hypotheses (normal and inverted neutrino mass ordering). When discussing the neutrino mass ordering at future neutrino oscillation experiments below, we will assume that the indicator used is the quantity typically referred to as the $\Delta\chi^2$:
\begin{equation}
 T = \Delta\chi^2 \equiv \min_{\rm IO} \chi^2 - \min_{\rm NO} \chi^2,
\end{equation}
where $\chi^2$ for given parameters is related to the likelihood $\mathcal L$ of the data according to $\chi^2 = -2\log(\mathcal L)$, thus making $T$ equivalent to the likelihood ratio of the best fit points in NO and IO. We will continue calling this quantity $T$, since it does typically not follow a $\chi^2$ distribution.

\section{Selecting between two hypotheses}
\label{sec:twomodels}

In particular, Bayesian model selection is very well suited for selecting between two mutually exclusive hypotheses where we are essentially certain that one of them has to be true. Such is the case of the neutrino mass ordering as was discussed by Qian et al.~\cite{Qian:2012zn}. In this section, we will largely follow their approach to analyze the Gaussian situation analytically in detail and also discuss the modifications introduced in a non-Gaussian scenario. We will later comment on the different possibilities of evaluating the merits future oscillation experiments based on Bayesian measures of sensitivity.

Let us first consider the general case, where the parameter space may be extended and the distribution of the indicator typically displays non-gaussianities regardless of the gaussianity of the underlying data. Once an experiment has been performed and resulted in a result $T = t$, we need to compute the posterior odds of the hypotheses $H$ and $\bar H$. By Bayes' theorem, this will be given by
\begin{equation}
 \frac{P(H;t)}{P(\bar H;t)} = \frac{\mathcal L_H (t)}{\mathcal L_{\bar H} (t)} \frac{\pi_H}{\pi_\barH} = \frac{p}{1-p}, \label{eq:postodds}
\end{equation}
where $p = p_{H}$ is the degree of belief in NO after making the observation $t$ and we have assumed $p_{H} + p_{\barH} = 1$.
Thus, in order to compute the posterior odds, compute the general integral in Eq.~(\ref{eq:likelihood}). In general, this may be a non-trivial task to perform analytically. However, it can be done numerically in a very straightforward fashion through simple Monte Carlo simulation 
as follows:
\begin{enumerate}
 \item \label{en:sample} If $H$ is a composite hypothesis, sample parameter values $\theta$ from the parameter prior $\pi_H(\theta)$.
 \item \label{en:data} Generate a set of data under the assumption that $H$ is true with parameters $\theta$.
 \item \label{en:compute} Compute and store the value of the indicator $T$ given the data generated in \ref{en:data}.
 \item \label{en:repeat} Repeat \ref{en:sample} to \ref{en:compute} until a sufficient number of samples of $T$ have been generated.
\end{enumerate}
This procedure will result in a sample of the distribution of $T$ under the assumption that $H$ is true. In some ways, this procedure is very similar to computing the distribution of a test statistic in a frequentist analysis. The big difference is that we are allowed to compute the distribution for $H$ as a whole rather than being restricted to doing it for a particular selection of parameters $\theta$. The reason we can do this is that the prior $\pi_H(\theta)$ has introduced a valid way of properly weighting the contributions from different parameter values. 

Once the likelihoods are known as a function of $t$ and the posterior odds computed, we can phrase our degree of belief in $H$ by using either the Kass--Raftery~\cite{KassRaftery} or Jeffrey~\cite{Jeffreys} scales. For comparison, it is useful to define the quantity
\begin{equation}
 \kappa(t) = 2 \log\left(\frac{P(H;t)}{P(\barH;t)} \right)
\end{equation}
 For the remainder of this text, we will use the Kass--Raftery scale, see Tab.~\ref{tab:KRscale}, but also give the degree of belief for reference.
\begin{table}
\begin{center}
\begin{tabular}{|l|c|c|c|}
\hline
{\bf Strength of evidence for $\boldsymbol H$} & $\boldsymbol \kappa$ & {\bf Posterior odds} & {\bf Degree of belief} \\
\hline
Barely worth mentioning & 0 to 2 & ca 1 to 3 & $< 73.11\%$\\
Positive & 2 to 6 & ca 3 to 20 & $> 73.11\%$ \\
Strong & 6 to 10 & ca 20 to 150 & $> 95.26\%$\\
Very strong & $> 10$ & $\gtrsim 150$ & $> 99.33\%$\\
\hline
\end{tabular}
\caption{The Kass--Raftery scale~\cite{KassRaftery} for wording the strength of evidence for a hypothesis $H$. \label{tab:KRscale}}
\end{center}
\end{table}

\subsection{Evaluating the performance of a future experiment}

The Bayesian framework gives us the freedom of computing a series of interesting probabilities already before an experiment has been performed. If we assume that $H$ and $\barH$ are the only two possible hypotheses which are equally probable before performing the experiment, then $P(H) = P(\barH) = 0.5$. As $\kappa(T)$ can be seen as the distribution of posterior odds once the experiment is performed. We can, for example, compute the probability of the experiment giving at least strong evidence~($>95.3\%$) for \emph{either} ordering
\begin{equation}
 P(|\kappa| > 6) = P(|\kappa| > 6; H) P(H) + P(|\kappa| > 6; \barH) P(\barH).
\end{equation}
This of course also includes the possibility of obtaining strong evidence for the wrong ordering (corresponding to a degree of belief $< 4.7\%$ for the correct ordering), so the more interesting quantity is
\begin{equation}
 P({\rm Strong\ correct}) = P(\kappa > 6; H) P(H) + P(\kappa < -6; \barH) P(\barH),
 \label{eq:strongcorrect}
\end{equation}
where we have exchanged the conditioned probabilities for obtaining strong evidence for the probability of obtaining strong evidence \emph{for the correct hypothesis}. Obviously $P({\rm Strong\ correct}) \leq P(|\kappa| > 6)$ as it should. Just as the sensitivity can be used to evaluate the performance of future experiments in the frequentist approach, probabilities such as this may be used to judge the merits of future experiments within a Bayesian approach. For example, we may judge experiments on the basis of their probability to give a certain level of evidence or on the evidence that the median experiment will provide. Below we will give examples of such performance indicators.

Note that Eq.~(\ref{eq:strongcorrect}) does not assume $H$ or $\barH$ to be true (as was done in Ref.~\cite{Qian:2012zn}), but is rather a weighted average of the results from both hypotheses with the weights given by the prior degree of belief in $H$ and $\barH$, respectively.
In the remainder of this paper, we will consider the situation where an equal prior probability of 0.5 has been assigned to both hypotheses. 
Furthermore, it should be pointed out that it is important to refrain from interpreting $\kappa$ or posterior degree of belief $p$ in terms of a number of $\sigma$, which is an inherently frequentist concept not present in a Bayesian analysis. The number of $\sigma$ at which a frequentist analysis would reject a hypothesis is dependent on the $p$-value, the probability of obtaining a result more extreme than the observed one if the hypothesis is true. In contrast, the $p$ computed in this Bayesian framework is the degree of belief that the hypothesis is true after an experiment is performed. These two concepts are fundamentally different and should not be confused.

\subsection{Normal distributed indicators}

In some limiting cases, the indicator $T$ will follow a normal distribution for both $H$ and $\barH$ with the same standard deviation.  
This limit is well fulfilled, e.g., for future reactor neutrino experiments sensitive to the neutrino mass ordering as discussed in Refs.~\cite{Qian:2012zn,Ciuffoli:2013rza}. For easy applicability to this scenario, we will assume that $T = N(\pm T_0,2\sqrt{T_0})$, where $N(\mu,\sigma)$ is a normal distribution with mean $\mu$ and standard deviation $\sigma$ and the $+$~($-$) is for NO~(IO). Clearly, any indicator which supplies normal distributions for $T$ can be brought to this form by translations and rescalings as long as they have the same standard deviation. For the particular case of the neutrino mass ordering at reactors, we also have $T_0 = \overline{\Delta\chi^2}$, i.e., the $\Delta\chi^2 = \min_\barH \chi^2 - \min_H \chi^2$ for the Asimov data set for $H$~\cite{Cowan:2010js}, i.e., the data set where all observables are given by their expectation values.

\begin{figure}
\begin{center}
\includegraphics[width=0.7\textwidth]{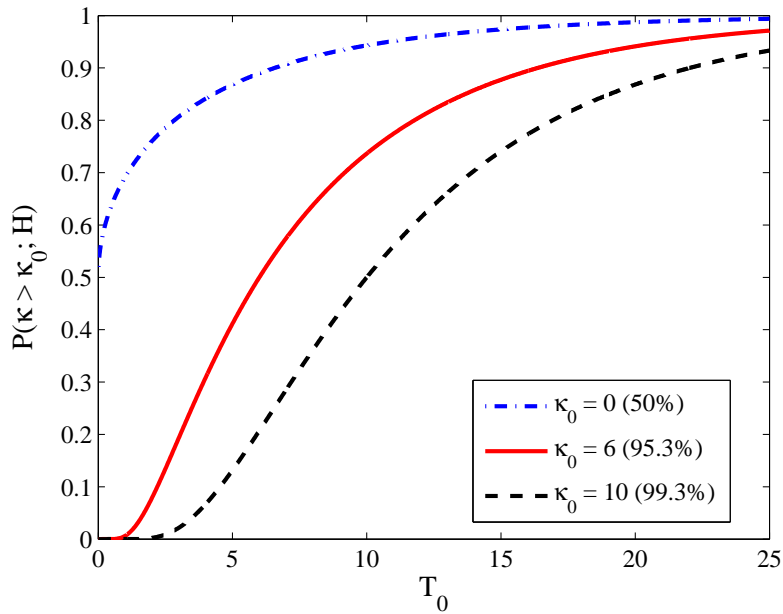}
\caption{The probability of obtaining strong or very strong evidence (corresponding to a posterior degree of belief of 95.3\% and 99.3\%, respectively, according to the Kass--Raftery scale) for the correct ordering in an experiment with Gaussian distributions for the indicator $T$ as a function of $T_0$. For reference, the dash-dotted blue curve shows the probability of actually favoring the correct ordering. \label{fig:evidenceprob}}
\end{center}
\end{figure}

Since the cumulative distribution and probability density functions of the normal distribution are known, we can obtain simple analytic expressions for several of the quantities and distributions mentioned above, in particular
\begin{eqnarray}
 \kappa(t) &=& t, \label{eq:kappagauss}\\
 P(\kappa > \kappa_0; H) &=& \frac 12 \left[1 - \erf\left(\frac{\kappa_0 - T_0}{2\sqrt {2T_0}} \right) \right], \label{eq:Pkappa} \\
 F(p) &=& \frac 12 \left[1 + \erf\left(\frac{2\log\frac{p}{1-p} - T_0}{2\sqrt{2T_0}}\right)\right],\\ 
 f(p) &=& \frac{1}{\sqrt{2\pi T_0}} \frac{1}{p(1-p)} \exp\left[ - \frac{\left(2\log\frac{p}{1-p} - T_0\right)^2}{8T_0} \right].
\end{eqnarray}
Here, 
$F(p)$ and $f(p)$ are the cumulative distribution and probability density functions for $p$ (before the experiment is performed, once an observation has been made $p$ is fixed) assuming that $H$ is true, respectively. Thus, by definition
\begin{equation}
 P(\kappa > \kappa_0; H) = 1 - F\left(\frac{1}{1+e^{-\kappa_0/2}}\right) = P(\kappa < -\kappa_0; \barH),
\end{equation}
where the last equality follows from symmetry, which agrees with Eq.~(\ref{eq:Pkappa}).
The probability density function $f(p)$ is the analytic form of what is shown in the left panel of Fig.~3 in Ref.~\cite{Qian:2012zn}.
Note that since the probability of obtaining evidence at least at level $\kappa_0$ for the true hypothesis in this case is independent of the hypothesis (this is true for all cases where the indicator is anti-symmetric under the exchange of $H$ and $\barH$) and thus
\begin{equation}
 P(\rm{evidence\ at\ least}\ \kappa_0\ {\rm for\ true\ hypothesis}) = \frac 12 \left[1 - \erf\left(\frac{ \kappa_0 - T_0}{2\sqrt {2T_0}} \right) \right].
 \label{eq:evidence}
\end{equation}
Note that this quantity is the probability of obtaining evidence at least at level $\kappa_0$ for the true hypothesis \emph{before} the experiment has been performed. Of course, once the experiment has been performed, we either have found the evidence or not. In particular, the quantity
\begin{equation}
 P(\kappa > 0; H) = \frac 12 \left[1 + \erf\left(\sqrt{\frac{T_0}{8}} \right)\right]
\end{equation}
is the probability that the experimental outcome will favor the correct ordering. A priori, this quantity has little to do with the actual posterior odds assigned to $H$ and $\barH$ as claimed in Ref.~\cite{Ciuffoli:2013rza}. If the experiment then measures $t = 0+\varepsilon$, the posterior odds will be close to one (equal degree of belief in $H$ and $\barH$) regardless of $T_0$. In other terms, the experiment did not provide us with significant discrimination between $H$ and $\barH$ although we might or might not have expected it to do so. This is neither strange nor unwanted. In fact, this is how we want the posterior odds to behave. If the results are essentially as likely in $H$ as in $\barH$, then a small shift in the experimental outcome should not significantly change our interpretation of the experiment.
Of more interest is the probability of actually obtaining strong (95.3\%) or very strong (99.3\%) evidence for the correct hypothesis, we show these probabilities as a function of $T_0$ in Fig.~\ref{fig:evidenceprob}.

To conclude this section, let us finally note that, since $\erf(0) = 0$, the median experiment will give evidence of strength $\kappa = T_0$. This can be understood from the facts that $\kappa(t) = t$ and $T$ is symmetric around $\pm T_0$.

\section{Example of a non-Gaussian situation}
\label{sec:nongaussian}

As mentioned earlier, the general case is typically not analytically solvable but instead we must apply numerical simulations in a fashion similar to that found in frequentist statistics, where one in general must simulate the distribution of the test statistic. However, contrary to the frequentist framework, the appearance of a probability measure on the parameters means that we will only need to perform one simulation for each hypothesis, rather than once for every parameter set. This detail in itself offers a significant reduction in the computing power needed to perform a test. In particular, the actual testing once the experiment has been performed turns into a simple matter of averaging the likelihood of the acquired data over each hypothesis to find the hypothesis likelihood and, through it, the posterior odds.

In this section, we will give an example of a situation where non-gaussianities play a role in the form of a simulation of the \NOvA\ experiment~\cite{Ayres:2004js}. In order to perform this analysis we will utilize the GLoBES software~\cite{Huber:2004ka,Huber:2007ji} along with selected parts of the MonteCUBES plug in~\cite{Blennow:2009pk}. Since we are mainly concerned with the statistical challenges, we simply use the predefined \NOvA\ glb files from the GLoBES homepage~\cite{GLOBEShome,Huber:2009cw} to describe the experiment. For the neutrino oscillation parameters and their priors we impose the conditions displayed in Tab.~\ref{tab:oscparams}.
\begin{table}
\begin{center}
\begin{tabular}{|c|c|c|}
\hline
{\bf Parameter} & {\bf Central value} & {\bf Prior} \\
\hline
$\theta_{23}$ & $45^\circ$ & Constant \\
$\theta_{12}$ & $33^\circ$ & Constant \\
$\sin^2(2\theta_{13})$ & 0.1 & Constant \\
$\Delta m_{21}^2$ & $7.9\cdot 10^{-5}$~eV$^2$ & Constant \\
$\Delta m_{31}^2$ & $\pm 2.4\cdot 10^{-3}$~eV$^2$ & Gaussian $\pm 10$\% \\
$\delta$ & -- & Flat, cyclic \\
\hline
\end{tabular}
\caption{Priors on the oscillation parameters used in our simulated non-Gaussian scenario studying the \NOvA\ experiment.\label{tab:oscparams}}
\end{center}
\end{table}
We only allow $\Delta m_{31}^2$ and $\delta$ to vary and fix the rest of the parameters. Again, this is done mainly for illustration purposes and in fact the deviation of $\theta_{23}$ from its maximal value is an additional challenge for future oscillation experiments, which we will comment on more later. In addition, note that we use these priors only for sampling the parameters in order to find the indicator distributions. We then follow the steps described in the previous section in order to compute these distributions.

\begin{figure}
 \begin{center}
  \includegraphics[width=0.7\textwidth]{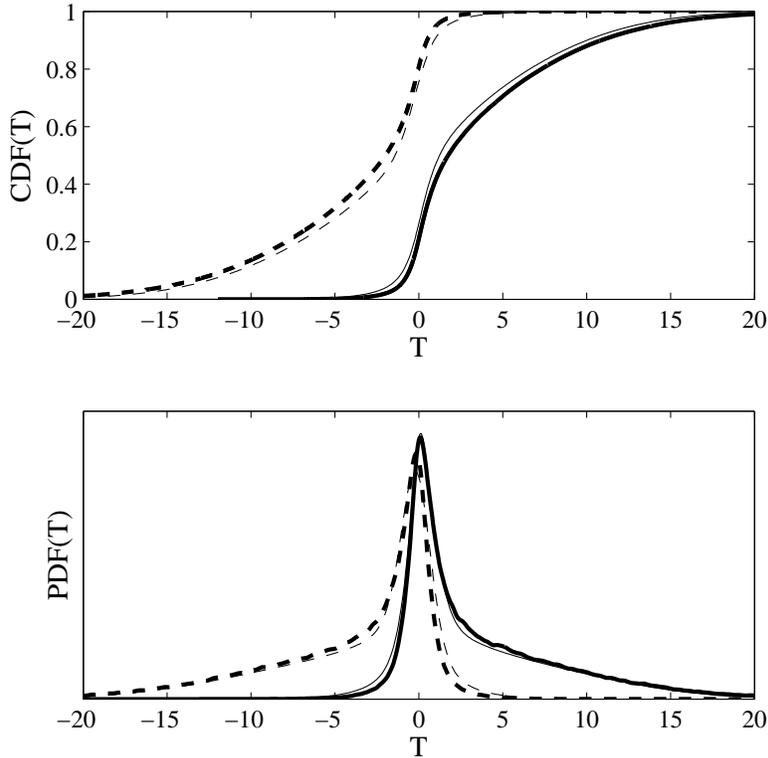}
  \caption{Cumulative distribution (top panel) and probability density (bottom panel) functions for $T_{\rm NO}$ (solid curves) and $T_{\rm IO}$ (dashed curves), respectively, in our simulation of the \NOvA\ experiment as described in the text. The non-gaussianities are clearly visible in asymmetry in the top panel and the large tails of the bottom panel. The thick curves correspond to the result of full simulation while the thin curves correspond to the distributions obtained in the semi-analytic Gaussian approximation (see Sect.~\ref{sec:semiGauss}).\label{fig:NOVAdf}}
 \end{center}
\end{figure}

\begin{figure}
\begin{center}
\includegraphics[width=0.7\textwidth]{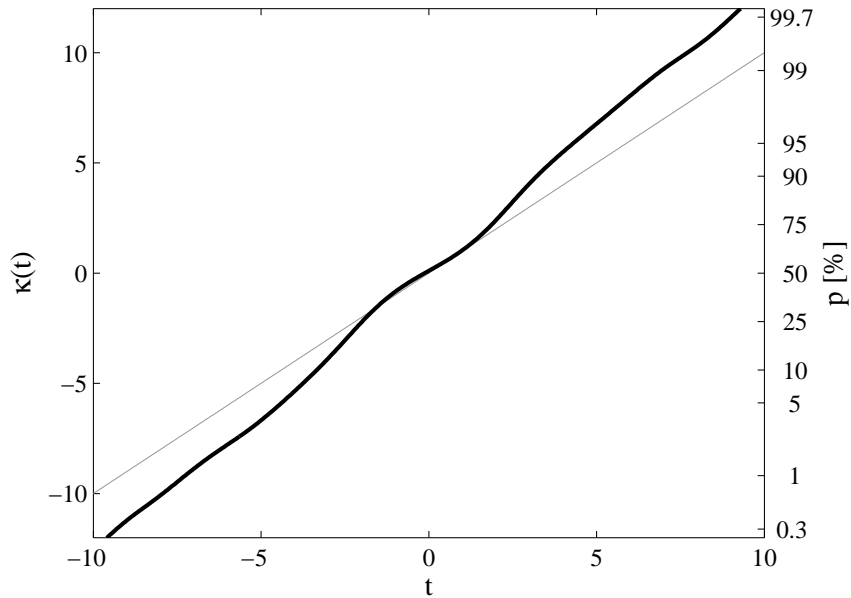}
\caption{The solid curve shows the value of $\kappa$ as a function of the measured value of the indicator $T = \min_{\rm IO}\chi^2 - \min_{\rm NO}\chi^2$ for the simulated \NOvA\ experiment. The thin gray curve shows the corresponding result for a Gaussian distribution. The scale on the right shows the corresponding posterior degree of belief $p$ in normal neutrino mass ordering.\label{fig:kappa}}
\end{center}
\end{figure}

In Fig.~\ref{fig:NOVAdf}, we show the cumulative distribution and probability density functions of the indicator $T = \min_{\rm IO} \chi^2 - \min_{\rm NO} \chi^2$ (we here use the built-in GLoBES $\chi^2$ function, which is $-2\log\mathcal L$ for a given parameter set).
As can be seen from the figure, the main difference from a Gaussian distribution is the appearance of large tail distributions in the direction of the simulated ordering. These tails arise mainly from regions of parameter space where the hierarchies are well separated. In this case, values of $\delta = \pm \pi/2$ are either close to the other ordering or the ones furthest removed from it. If the value of $\delta$ is such that it will be easy to separate the hierarchies, then we have a significant chance of obtaining evidence for the true ordering. However, this is where the prior on the parameter space of the models come into play and properly weights the probabilities of being in different regions of parameter space. This is in some sense similar to the gain fraction argument for CP-violation presented in Ref.~\cite{Blennow:2013swa}, with the difference that we are now performing a full Bayesian analysis. From the probability density functions, it is now easy to construct the value of $\kappa(t)$ and we show this quantity in Fig.~\ref{fig:kappa}.
As can be seen in this figure, the posterior odds for a given value of the indicator $T$ will typically be larger than the Gaussian $t$. This effect is due to the tails of the probability density functions. Since $\kappa$ is related to the ratio of the probability density functions and these fall off much slower than in the Gaussian case on the side of the true ordering, it follows that this will generally be the case. If instead we had a situation where the tails were on the other side, then we would typically obtain weaker posterior odds for the same $t$.

It should be stressed that the results shown in Fig.~\ref{fig:kappa} is the only important piece of information once the experiment has been performed as relating the measured value of the indicator with the posterior odds is straightforward using this figure. As expected, a measured value around $t = 0$ adds very little information on whether NO or IO is the true one.

\begin{figure}
\begin{center}
\includegraphics[width=0.65\textwidth]{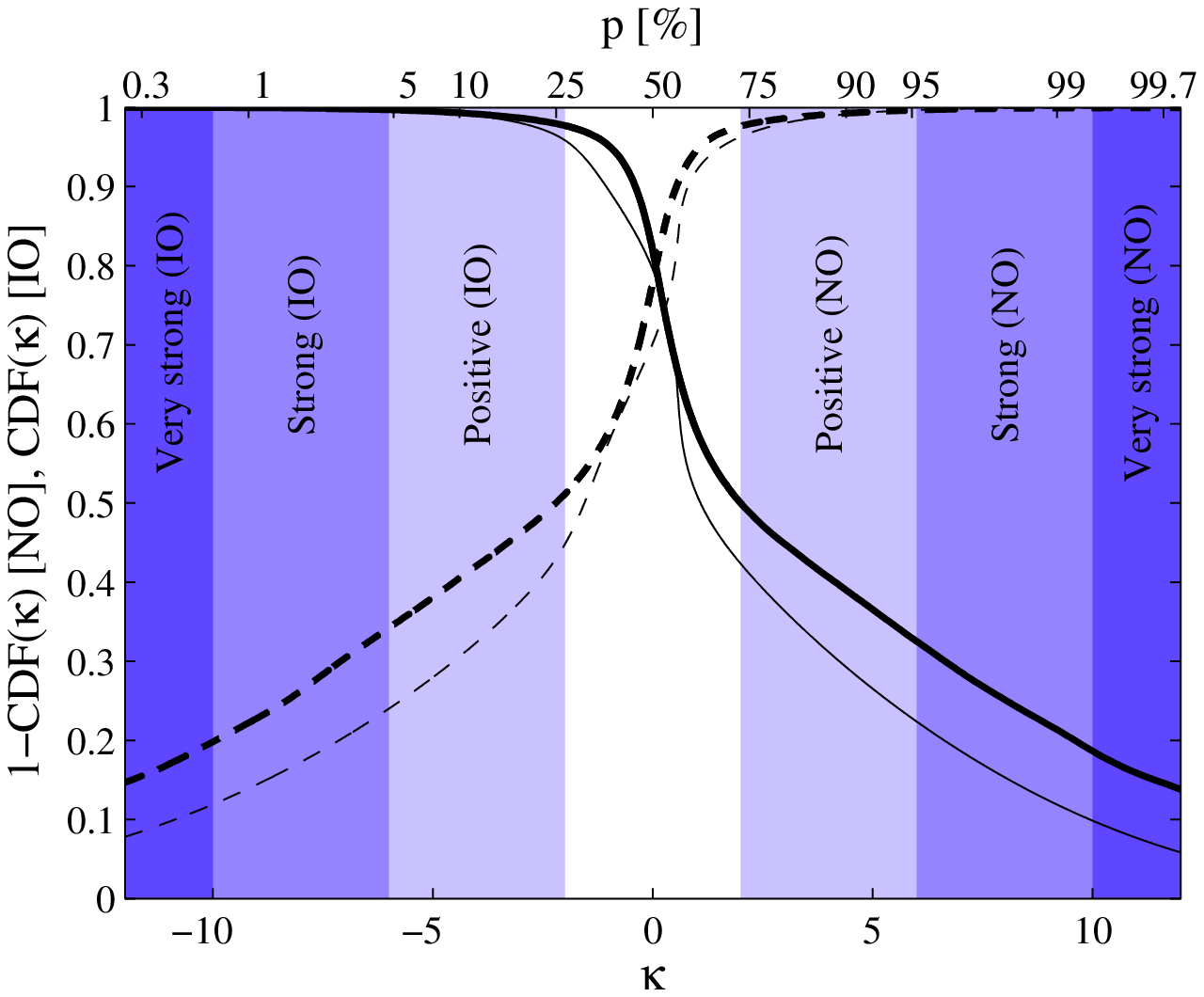} \\\includegraphics[width=0.65\textwidth]{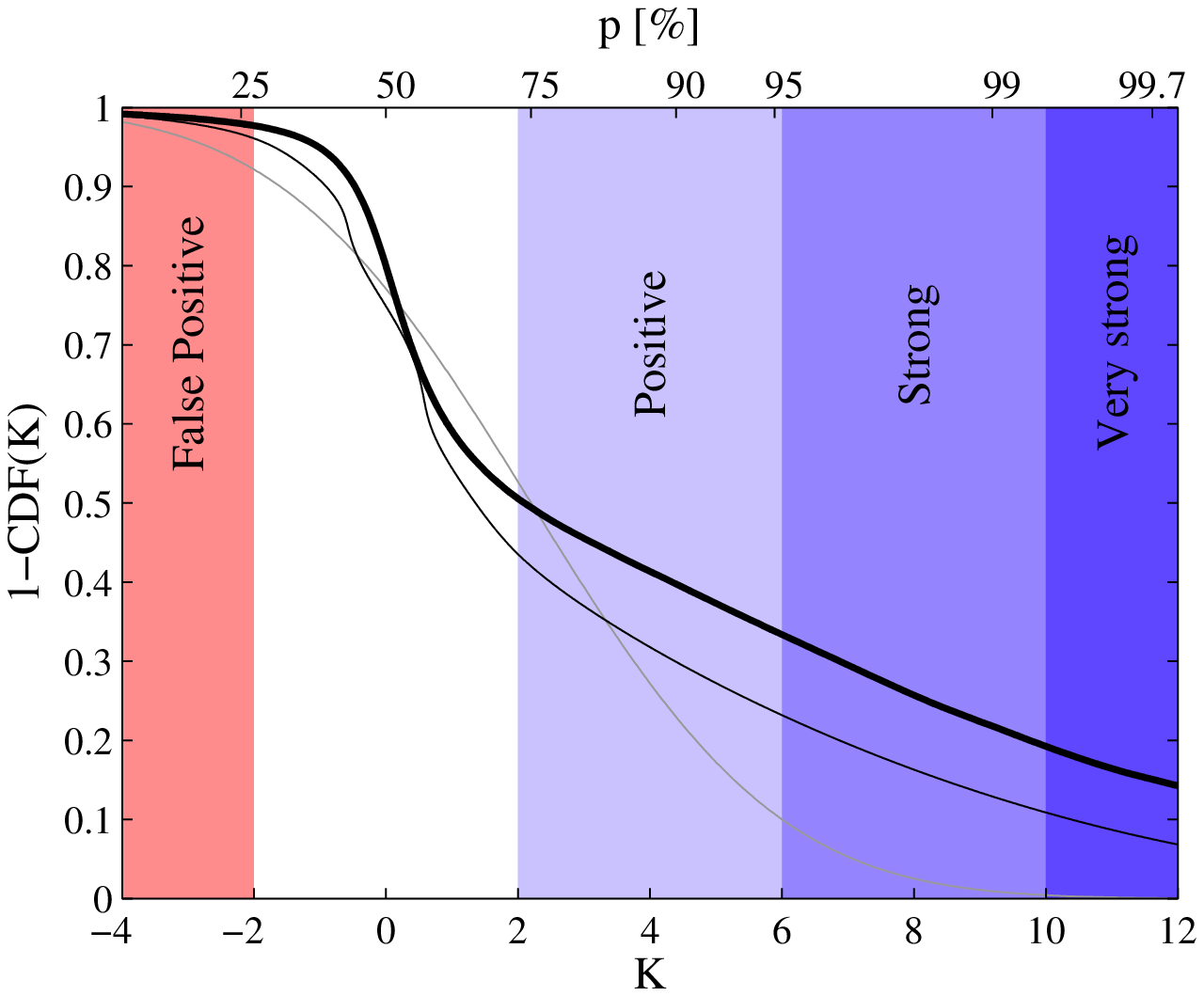}
\caption{Top panel: Probability of obtaining posterior odds corresponding to $\kappa$ \emph{or higher} in favor of the correct ordering. The solid curve assumes NO, while the dashed curve assumes IO. Note that the curves are tilted in opposite directions as IO becomes more likely as the value of $\kappa$ decreases. Bottom panel: The probability of obtaining posterior odds corresponding to $K$ \emph{or higher} in favor of the correct ordering. We use $K = \kappa$ for true NO and $K = - \kappa$ for true IO in order to collect the probability for obtaining evidence for the true ordering. As reference, the thin gray curve shows the corresponding quantity for a Gaussian experiment with $T_0 = 2.2$, which gives the same median posterior odds. The scale on the top of the upper (lower) panel show the corresponding posterior degree of belief in the normal (correct) neutrino mass ordering. The thin black curves show the corresponding results of the semi-analytic approximation presented in Sec.~\ref{sec:semiGauss}.\label{fig:kappacdf}}
\end{center}
\end{figure}

Finally, we will now discuss how to judge the capabilities of a future experiment within the Bayesian framework and argue that the full distribution of the expected evidence should be considered rather than reducing it to a single number. This will roughly correspond to the typical sensitivity analysis that is usually performed in the frequentist setting. In order to do this, we compute the distributions of $\kappa$ under the assumptions of NO and IO. This is done by simply inserting the indicator distributions for NO and IO, respectively, and thus obtaining the corresponding distributions of $\kappa$. These distributions, whose cumulative distribution functions are shown in Fig.~\ref{fig:kappacdf}, give us an estimate of what we can expect for the outcome of the experiment. 
For example, we see that $P(\kappa > 10; {\rm NO}) = 1-\CDF_{\kappa,{\rm NO}}(10) \simeq 0.2$. Thus, the probability of \NOvA\ presenting very strong evidence ($>99.3\%$) for NO if NO is true is around 20\%. Since the situation is relatively symmetric, the same is true for IO and thus the probability of \NOvA\ presenting very strong evidence for the true ordering is also around 20\%. Also note that the probability of presenting very strong evidence for the false ordering ($<0.7\%$ degree of belief in the correct ordering) is miniscule. In order to compare different possibilities for future experiments, one can either construct the equivalent of Fig.~\ref{fig:kappacdf} for all experiments and use them to judge the pros and cons of the different experiments. It may be possible that experiment A provides a greater chance of obtaining at least strong evidence, while experiment B provides a greater chance of obtaining at least very strong evidence. The alternative is to pick a desired level of evidence and just compare the experiments' probabilities of reaching at least that level. While the disadvantage of this procedure is a certain loss of information, the advantage is that this single number can be plotted against, e.g., experimental parameters such as the neutrino energy, baseline, or running time. We wish to stress the fact that, since the shape of the distribution of $\kappa$ in general will vary depending on the experiment, a single number will typically be insufficient to describe the full situation. For example, a Gaussian experiment with $T_0 = 4$ would have a larger median evidence for the true ordering than our simulated \NOvA\ experiment. However, the chances of getting lucky and obtain very strong evidence in such an experiment would be miniscule. Thus, rather than tabulating several predefined numbers describing the distributions, we would advocate simply showing the distribution and basing any comparison on how the distributions compare.

\subsection{Semi-analytic approximation}
\label{sec:semiGauss}

It was observed in Ref.~\cite{Blennow:2013oma}, that for fixed values of the oscillation parameters (in particular $\delta$), the distribution of $T$ would still take on a Gaussian form with mean and standard deviation now given by the $T_0(\theta)$ produced by that particular parameter set $\theta$. If this holds and $T_0(\theta)$ is known, then the probability density function of $T$ will be given by
\begin{equation}
 \Li_{\rm NO}(T) = \int \pi_{\rm NO}(\theta) \Li_{T_0(\theta)}(T) d\theta,
\end{equation}
where
\begin{equation}
 \Li_{T_0}(t) = \frac{1}{\sqrt{8\pi T_0}} \exp\left(-\frac{(t-T_0)^2}{8T_0}\right)
\end{equation}
is the probability density function of $N(T_0,2\sqrt{T_0})$ and the corresponding expressions hold for inverted ordering. If $T_0(\theta)$ has the same distribution in both hierarchies, i.e.,
\begin{equation}
 \pi_{\rm NO}(\theta) \frac{d \theta}{dT_0} dT_0 = \pi_{\rm IO}(\bar\theta) \frac{d\bar\theta}{d T_0} dT_0 \equiv \pi(T_0) dT_0 
\end{equation}
for all $T_0$, then
\begin{eqnarray}
 \Li_{\rm NO}(t) = \exp\left(\frac t4 \right)\int \pi(T_0) \frac{1}{\sqrt{8\pi T_0}} \exp\left( - \frac{t^2 + T_0^2}{8T_0^2} \right) dT_0, \\
 \Li_{\rm IO}(t) =  \exp\left(-\frac t4 \right)\int \pi(T_0) \frac{1}{\sqrt{8\pi T_0}} \exp\left( - \frac{t^2 + T_0^2}{8T_0^2} \right) dT_0.
\end{eqnarray}
It follows directly that also in this case,
\begin{equation}
 \kappa(t) = 2\log\left(\frac{\Li_{\rm NO}(t)}{\Li_{\rm IO}(t)}\right) = t.
\end{equation}
This is a generalization of the result found in Eq.~(\ref{eq:kappagauss}) and any deviations from this result must be caused by non-Gaussianities or different distributions of $T_0(\theta)$ in the different orderings.

The largest deviation from this appeared for the case of the \NOvA\ experiment, which is the example we have been using so far in this work. For comparison, we show the expectation of this semi-analytic approximation with thin curves in the lower panel of Fig.~\ref{fig:NOVAdf}. As can be seen, the approximation overestimates the value of the probability density function where the ordering is not the assumed one, while it underestimates it in the other tail. For producing this result, we have computed $T_0$ as a function of $\delta$ for both hierarchies independently and the results do not provide the same $T_0$ distributions although it is close enough for $\kappa(t)$ not to show any significant deviations from $t$ for $|t| < 20$. Thus, the deviations from this rule seen in Fig.~\ref{fig:kappa} originate mainly from the non-Gaussianity of the \NOvA\ distributions. Thus, using this semi-analytic approach for \NOvA\ will therefore lead to conservative estimates of the \NOvA\ capabilities. We also show the result of the approximation as thin black curves in Fig.~\ref{fig:kappacdf}. This figure confirms our expectation that the approximation will underestimate the capability of \NOvA, although the main features such as the larger tail probability are still present, the probability of reaching a good degree of belief may be underestimated by as much as 10\%.

\subsection{Prior dependence}

Just as with any Bayesian statement, the distribution of the posterior odds prior to an experiment is performed will typically be dependent on the parameter prior within the models. In the situation described above, the main prior impact arises from the flat prior on $\delta$, while the actual prior chosen for $\Delta m_{31}^2$ does not change the prediction significantly. This mainly occurs since the probability of generating data that will separate the hierarchies is highly $\delta$ dependent. This only reflects the fact that when considering what future experiment to built, the current knowledge of the model parameters should also be taken into account (see also Ref.~\cite{Blennow:2013swa}). For the CP-violating phase, most people would agree that a flat prior is a reasonable starting point as long as no data is available to favor one region of $\delta$ or the other. However, once there is a hint for a value of $\delta$, it is only natural that the perceived probability of making a given measurement should change accordingly.

The prior dependence would play an even bigger role in a situation where there is no lower bound on the effect that is being searched for. In the case of the mass ordering, $\Delta m_{31}^2 = 0$ has long since been ruled out. However, for the determination of the octant of $\theta_{23}$, the data of global fits are still very compatible with maximal mixing. Of course, the possibility to discriminate between the octants is crucially dependent on how far from maximal $\theta_{23}$ is allowed to be and a prior which allows larger deviations will typically give a more optimistic prediction for the capabilities of an experiment to discover the octant. Let us also note that if the parameter knowledge is not very precise, leading to good perceived chances of a discovery, and a discovery is later not made once the experiment is performed, the experiment should still typically not be considered a failure as it then will tend to disfavor those parts of parameter space where a discovery would be easy, thus providing a better prior as input for the next experiment.

\begin{figure}
\begin{center}
\includegraphics[width=0.7\textwidth]{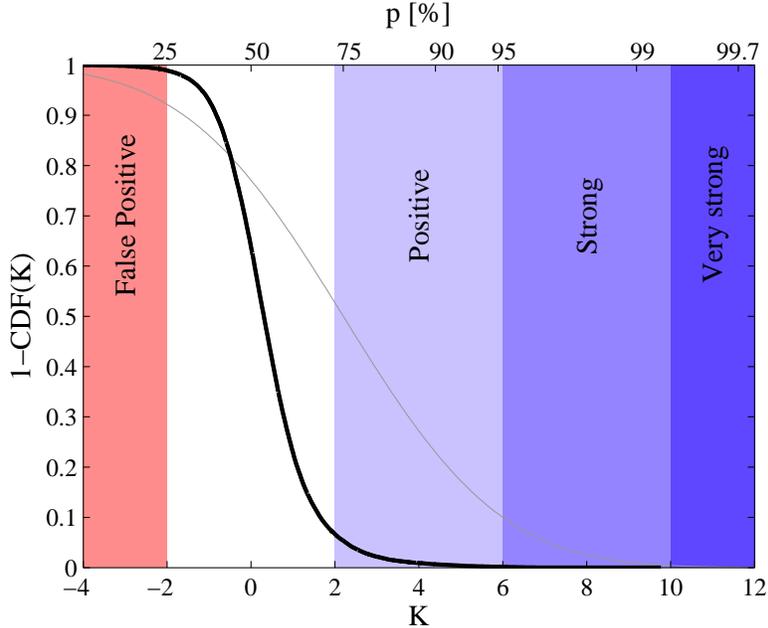}
\caption{The same as the lower panel of Fig.~\ref{fig:kappacdf}, but with a prior that fixes $\delta$ to the most pessimistic values for the purposes of mass ordering determination. \label{fig:priordependence}}
\end{center}
\end{figure}

As an example of the dependence on the prior, we show the case where $\delta$ is fixed to $90^\circ$ in NO and $-90^\circ$ in IO in Fig.~\ref{fig:priordependence}. Due to these points being almost degenerate, essentially no chance remains for obtaining any level of reliable evidence for which ordering is the true one through the \NOvA\ experiment.

\section{Summary and discussion}
\label{sec:summary}

We have discussed and given concrete examples of how to apply a Bayesian analysis to the problem of neutrino mass ordering determination. In doing so, we have described the correct procedures to obtain the Bayesian posterior odds for any experiment and discussed how Bayesian methods may be used to judge the capabilities of future neutrino oscillation experiments by extending the approach and argumentation presented by Qian et al.~\cite{Qian:2012zn}. In short, before an experiment is performed, one may speak only about the expected distribution of the posterior odds of the two different orderings. Here, the actual value of the indicator, typically taken to be $\Delta\chi^2 = \min_{\rm IO}\chi^2 - \min_{\rm NO}\chi^2$, and not only its sign is of importance. We want to point out that this is precisely what we would expect in a Bayesian setting, as changing a measurement by a small amount from $\Delta\chi^2 = \varepsilon$ to $-\varepsilon$ should not change the results significantly. We contrast this to the situation presented by Ciuffoli et al.~Ref.~\cite{Ciuffoli:2013rza}, where the ordering is identified as normal if $\Delta\chi^2 > 0$ and the probability of obtaining $\Delta\chi^2 > 0$ is taken as the confidence level. This corresponds to attempting to determine the confidence level in a frequentist fashion, i.e., asking the question of how large the ratio of infinitely repeated experiments would arrive at the correct ordering. In the frequentist nomenclature, this is simply a computation of the confidence level provided by a test which takes zero as the critical value. As we have shown, in the symmetric Gaussian approximation, this probability only corresponds to the chance of the correct ordering being favored after the experiment is performed and says nothing about how favored it has to be (just that it has to be above 50\%). A more interesting question than this would be to ask for the probability that an experiment will give at least strong or very strong evidence for the correct ordering, corresponding to a posterior degree of belief of at least $95.3\%$ and $99.3\%$, respectively, which we have also computed for the case of a Gaussian distribution of the indicator.

The second question of Ref.~\cite{Ciuffoli:2013rza} regards the confidence level obtained by a typical experiment. This is somewhat misguided, since confidence is an inherent frequentist concept and what is interesting in the Bayesian framework is simply the posterior odds. In particular, what is interpreted in Ref.~\cite{Ciuffoli:2013rza} as the probability of success is actually the posterior probability of the correct ordering, which the authors then go on to interpret as a frequentist $p$-value. While the $p$-value in a frequentist setting is the probability of obtaining a more extreme result, the posterior probability evaluated for the median experiment is the degree of belief in the correct ordering if the data would turn out to be the median expected data. Here we have also argued against the use of a number of $\sigma$ for describing the strength of the evidence, since in the frequentist nomenclature the neutrino community is familiar with, this represents a measure of how much the data deviates from the expectation and not the degree of belief in a model. The third question raised in Ref.~\cite{Ciuffoli:2013rza} is asking what the probability of achieving a certain level of evidence (although it is referred to as confidence), we have discussed this extensively throughout this text and the main results include Eq.~(\ref{eq:evidence}) for the Gaussian approximation and Fig.~\ref{fig:kappacdf} for the simulation of \NOvA.

In addition to the above, we have given examples of how to perform the full Bayesian analysis both analytically for the case where a Gaussian distribution can be assumed as well as numerically in the case of non-Gaussian distributions. Non-Gaussian distributions will typically result from appearance experiments where the value of the CP-violating phase $\delta$ is of importance. We have illustrated this by performing a simplified analysis of what may be expected from the \NOvA\ experiment and seen that the appearance of large tails in the distribution functions imply that values of $\Delta\chi^2$ typically give stronger evidence than for the Gaussian case. For the case of composite hypotheses composed of simple hypotheses with Gaussian distributions of the indicator, we have also provided a semi-analytic approach to slightly simplify the computation of the indicator distributions for the full hypothesis.
In the very end, we discussed the impact of priors in the Bayesian analysis and argued that the prior dependence is not necessarily a bad thing when it comes to decision making for future experiments, since their capabilities should be based on our current knowledge and beliefs about the true values of the underlying physics.

Let us finally remark that this analysis by no means is specific for the neutrino mass ordering. It is equally appropriate also for other binary measurements such as which octant $\theta_{23}$ belongs to, or any other binary (or even larger degeneracies) measurement in physics. In particular, it should be expected that the octant degeneracy of $\theta_{23}$ displays even more non-Gaussian behavior since, unlike for the neutrino mass ordering, the degenerate solutions are very weakly separated.

\begin{acknowledgments}
The author would like to thank P.~Coloma, P.~Huber, T.~Schwetz, and E.~Fernandez-Martinez for useful discussions during the completion of this work as well as S.~Zhou for useful comments.

This work was supported by the G\"oran Gustafsson Foundation.
\end{acknowledgments}

\end{document}